\documentclass[10pt,conference]{IEEEtran}
\IEEEoverridecommandlockouts
\usepackage{cite}
\usepackage{amsmath,amssymb,amsfonts}
\usepackage{algorithmic}
\usepackage{graphicx}
\usepackage{textcomp}
\usepackage{xcolor}

\usepackage{url}

\usepackage{booktabs} 
\usepackage{tabularx}
\usepackage{graphicx}
\usepackage{caption}
\usepackage{subcaption}
\usepackage{amsmath}
\usepackage{balance} 
\usepackage{amssymb}
\usepackage{xspace}
\usepackage[T1]{fontenc}
\usepackage[scaled=0.81]{beramono}
\usepackage{balance}

\usepackage{enumitem}  
\usepackage{listings}
\usepackage{url}
\usepackage[ruled,lined,linesnumbered,vlined]{algorithm2e}
\usepackage{multirow}
\usepackage{array}
\usepackage{color}
\usepackage{float}
\usepackage[misc]{ifsym}

\usepackage{listings, xcolor}

\definecolor{verylightgray}{rgb}{.97,.97,.97}

\lstdefinelanguage{Solidity}{
	keywords=[1]{anonymous, let, assembly, assert, balance, break, call, callcode, case, catch, class, constant, continue, constructor, contract, debugger, default, delegatecall, delete, do, else, emit, event, experimental, export, external, false, finally, for, function, gas, if, implements, import, in, indexed, instanceof, interface, internal, is, length, library, log0, log1, log2, log3, log4, memory, modifier, new, payable, pragma, private, protected, public, pure, push, require, return, returns, revert, selfdestruct, send, solidity, storage, struct, suicide, super, switch, then, this, throw, transfer, true, try, typeof, using, value, view, while, with, addmod, ecrecover, keccak256, mulmod, ripemd160, sha256, sha3}, 
	keywordstyle=[1]\color{blue}\bfseries,
	keywords=[2]{address, bool, byte, bytes, bytes1, bytes2, bytes3, bytes4, bytes5, bytes6, bytes7, bytes8, bytes9, bytes10, bytes11, bytes12, bytes13, bytes14, bytes15, bytes16, bytes17, bytes18, bytes19, bytes20, bytes21, bytes22, bytes23, bytes24, bytes25, bytes26, bytes27, bytes28, bytes29, bytes30, bytes31, bytes32, enum, int, int8, int16, int24, int32, int40, int48, int56, int64, int72, int80, int88, int96, int104, int112, int120, int128, int136, int144, int152, int160, int168, int176, int184, int192, int200, int208, int216, int224, int232, int240, int248, int256, mapping, string, uint, uint8, uint16, uint24, uint32, uint40, uint48, uint56, uint64, uint72, uint80, uint88, uint96, uint104, uint112, uint120, uint128, uint136, uint144, uint152, uint160, uint168, uint176, uint184, uint192, uint200, uint208, uint216, uint224, uint232, uint240, uint248, uint256, var, void, ether, finney, szabo, wei, days, hours, minutes, seconds, weeks, years, u64},	
	keywordstyle=[2]\color{teal}\bfseries,
	keywords=[3]{block, blockhash, coinbase, difficulty, gaslimit, number, timestamp, msg, data, gas, sender, sig, value, now, tx, gasprice, origin},	
	keywordstyle=[3]\color{violet}\bfseries,
	identifierstyle=\color{black},
	sensitive=false,
	comment=[l]{//},
	morecomment=[s]{/*}{*/},
	commentstyle=\color{gray}\ttfamily,
	stringstyle=\color{red}\ttfamily,
	morestring=[b]',
	morestring=[b]"
}

\newcommand{\etal}{{\emph{et al.}} }

\newcommand{\ie}{{\emph{i.e.}} }

\def\BibTeX{{\rm B\kern-.05em{\sc i\kern-.025em b}\kern-.08em
		T\kern-.1667em\lower.7ex\hbox{E}\kern-.125emX}}
\begin{document}
	
\title{Performance Analysis of the Libra Blockchain: An Experimental Study}

\author{
	\IEEEauthorblockN{
		Jiashuo Zhang\IEEEauthorrefmark{1}, 
		Jianbo Gao\IEEEauthorrefmark{1}\IEEEauthorrefmark{2}, 
		Zhenhao Wu\IEEEauthorrefmark{1}
		Wentian Yan\IEEEauthorrefmark{1}, 
		Qize Wu\IEEEauthorrefmark{2}, 
		Qingshan Li\IEEEauthorrefmark{1}\IEEEauthorrefmark{2},
		Zhong Chen\IEEEauthorrefmark{1}\IEEEauthorrefmark{3}
	}
	\IEEEauthorblockA{\IEEEauthorrefmark{1}\textit{School of Electronics Engineering and Computer Science}, \textit{Peking University}, Beijing, China} 
	\IEEEauthorblockA{\IEEEauthorrefmark{2}\textit{Boya Blockchain Inc.}, Beijing, China}
	\IEEEauthorblockA{\IEEEauthorrefmark{3}Corresponding Author}
	\{zhangjiashuo, gaojianbo, zhenhaowu, yanwentian2018\}@pku.edu.cn, wuqz@boyachain.cn, \{liqs, zhongchen\}@pku.edu.cn

}

\maketitle

\begin{abstract}
	Since Bitcoin was first introduced in 2008, many types of cryptocurrencies have been proposed based on blockchain.
However, the performance of permissionless blockchains restricts the widespread of cryptocurrency.
Recently, Libra was proposed by Facebook based on a permissioned blockchain, \ie the Libra blockchain.
The vision of Libra is to become a global currency supporting financial applications, but it is doubted whether the performance of the Libra blockchain is able to support frequent micropayment scenarios. 
In this paper, we propose a methodology to evaluate the performance of blockchain platforms and conducted an experimental study on the Libra blockchain.
The results show that the Libra blockchain can only process about one thousand transactions per second at most, and the performance drops significantly as the number of validators increases.
Although it outperforms permissionless blockchain platforms, the performance of the Libra blockchain is still unsatisfactory compared to other permissioned blockchains like Hyperledger Fabric and needs to make effective improvements in order to support global micropayment in the future.

\end{abstract}

\begin{IEEEkeywords}
	Performance Analysis, Blockchain, Libra, Hyperledger Fabric

\end{IEEEkeywords}

\section{Introduction}
\label{sec:intro}
Blockchain has become popular all over the world since it was first introduced by Bitcoin\cite{nakamoto2008bitcoin} in 2008. 
More and more blockchain platforms with different characteristics are proposed to meet different demands. 
Due to its advantages like decentralization, non-modifiability, and security, it has become the core technology of cryptocurrency systems and has been used to many other areas like finance and supply chain.

According to allowing anonymous public nodes or not, blockchain can be classified into two types, \ie permissioned and permissionless blockchains \cite{vukolic2017rethinking}. There are some trade-offs between them and the most important metrics to be considered are performance and scalability.
Permissionless blockchains, represented by Bitcoin and Ethereum \cite{wood2014ethereum}, usually use PoW \cite{garay2015bitcoin} as their consensus protocol which causes high latency and low throughput. Their Tps is often below 100 while Visa's Tps is about 2000 \cite{Lee:2019:BPR:3307334.3328666} which makes them difficult to become global economic infrastructures. 
Permissioned blockchains represented by Hyperledger Fabric \cite{Androulaki:2018:HFD:3190508.3190538} usually have better performance because their consensus protocols, such as PBFT \cite{castro1999practical}, are more efficient than PoW \cite{bach2018comparative}. But their scalability become a new bottleneck. Considering performance is more important than scalability in practical usage, the permissioned blockchains are still the most promising candidate to support global application scenarios \cite{lewis2017blockchain}.

Libra\cite{Amsden2019} is introduced by Facebook recently, and it has attracted
worldwide attention. It is a permissioned blockchain aiming to provide 
global service as a common currency. 
However, it still faces the same question like other permissioned blockchains. To solve that, 
it implemented many new properties to support high performance and scalability. 
This paper focuses on these new properties and evaluates them with scientific data support to help the community understand Libra better.
The contributions of this paper are summarized as follows:
\begin{itemize}
    \item We proposed a methodology to evaluate the blockchains with batch operations and automatic evaluation. We also conduct a tool with this methodology for Hypledger Fabric and Libra, 
    which can support different numbers of clients to submit transactions concurrently.
    \item We evaluate the performance and scalability of Libra. It is a novel work and nobody has finished it as we know. We also discuss Libra
    from the possibility of being a global currency.
    \item We explore the influence of different layers in Libra, which can help the community understand the performance of the whole platform, find the bottleneck and improve it.
\end{itemize}

The remaining of this paper is organized as follows: Section~\ref{sec:background} provides a brief introduction of Hypledger Fabric and Libra. In Section~\ref{sec:methodology}, we explain the methodology we used in our experiment. In Section~\ref{sec:results}, we present our results and analyse them. In Section~\ref{sec:relatedwork}, we discuss some relevant previous work. Finally, Section~\ref{sec:conclusion} concludes this paper.

\section{Background}
\label{sec:background}
\begin{figure*}
	\centering
	\includegraphics[width=0.8\linewidth]{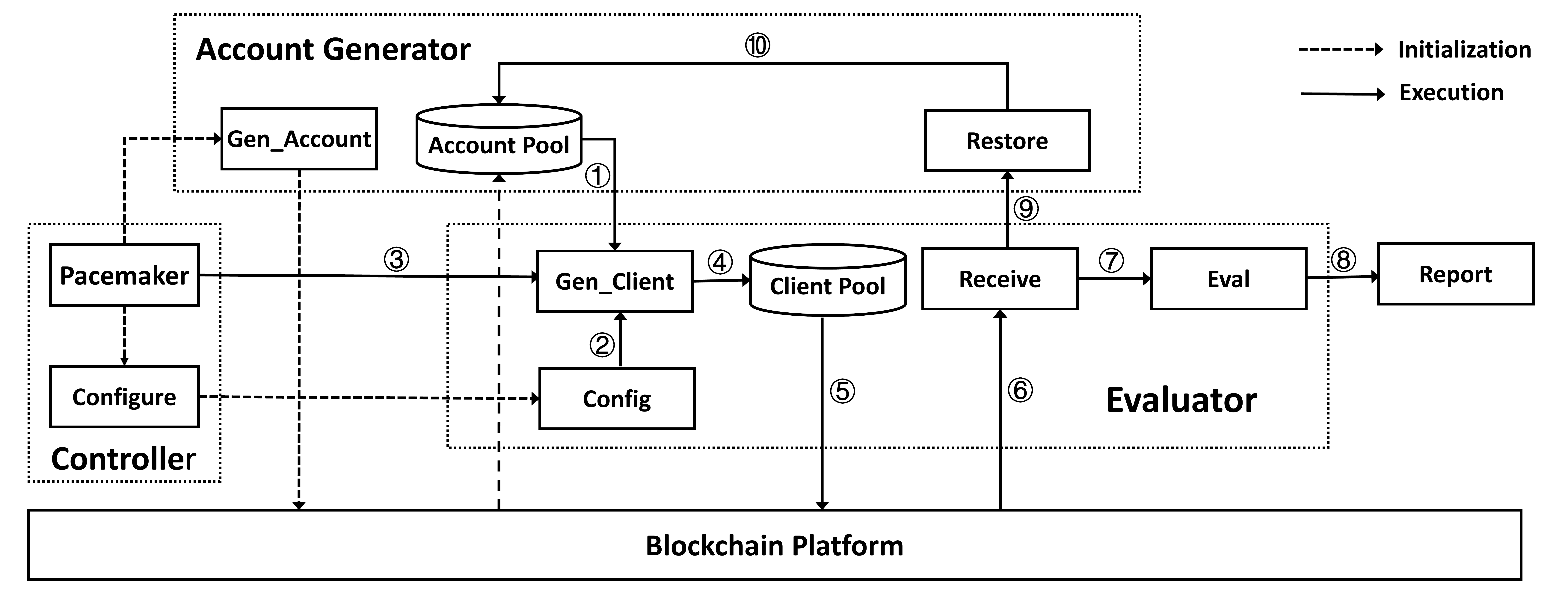}
	\caption{Overall workflow of our tools, which mainly includes the account generator for account generation and minting, evaluator for account loading and evaluating, and controller to support batch evaluation. Dotted arrows represent the workflow during the initialization phase and solid arrows represent the workflow during the execution phase.}
	\label{fig:overview}
	
\end{figure*}

In this section, we briefly introduce the target platforms we choose in our experiment, \ie Hyperledger Fabric and Libra. As the main difference between different blockchains are consensus protocols and the execution of smart contracts, we mainly introduce Hyperledger Fabric and Libra from these two aspects.


Hyperledger Fabric is an open-source, permissioned and distributed ledger technology platform with highly modular and configurable architecture. Due to its high flexibility, many different applications are developed from it to solve various scenarios. Hyperledger Fabric supports pluggable consensus protocols such as PBFT in Fabric v0.6 and Raft \cite{ongaro2014search} in Fabric v1.4.1. 
The smart contract of Fabric is called chaincode, through which users can implement business logic and interact with the ledger.
In Fabric v1.4.1, the consensus protocol is improved by a kind of new nodes called orderers, which can improve the scalability and performance of Fabric. 
Benefit from the introduction of orderers, Hyperledger Fabric v1.4.1 becomes a more powerful platform, and that makes us choose it to be the benchmark compared with Libra.

The Libra blockchain \cite{Amsden2019} is an open-source, permissioned blockchain aiming to become a global currency and create a new opportunity for financial services innovation. 
The value of Libra is guaranteed by The Libra Association which is comprised of many reliable members such as businesses, nonprofit and multilateral organizations.  
It has already conducted a healthy and stable financial ecosystem which will support Libra to become faithful. 

Move\cite{Blackshear2019} is a safe and flexible programming language designed for Libra blockchain. The key feature of it is the first-class resources. 
These resources obey the linear logic so that they can not be copied or implicitly discarded, and the only valid operation is moving them between program storage locations. 
Each bytecode will be checked by a module called bytecode verifier before execution to ensure it obeys the linear logic and other security restrictions. Move guarantees the safety of executing transaction scripts in Libra. Libra uses a VM called Move VM to execute Move scripts and the quality of VM can influence the performance of the whole system.

LibraBFT\cite{Baudet2019} is a robust and efficient state machine replication (SMR) system designed for the Libra blockchain. It is a variant of BTF based on Hotstuff\cite{yin2019hotstuff} and focuses on stronger scalability. The participants of SMR are called validators 
which are responsible for handling transactions sent by clients and maintaining consensus on the history of transactions among honest validators. For consistency and simplicity, we call validators ``peers'' in the remaining of this paper.
LibraBFT is designed for strong scalability while other consortium blockchains like Hyperledger Fabric v0.6 may only support less than 30 peers \cite{nasir2018performance}. The early prototypes of LibraBFT also showed a promising performance in throughput and latency which will support large scale applications with efficiency requirement.

\section{Methodology}
\label{sec:methodology}
\subsection{Experimental Design}
 
Our tools are divided into three parts as shown in Fig.~\ref{fig:overview}, \ie Controller, Account Generator, and Evaluator. Controller consists of Pacemaker to automatically control the workflow and Configure modular to configure parameters, such as the number of transactions.  Account Generator is responsible for creating accounts, storing them in the pool and providing them to the Evaluator. Evaluator pulls accounts from the pool and uses them to evaluate. 
Both of Evaluator and Account Generator follow the instruction from the Controller.  We conduct our tools both on Libra and Fabric and all of the data below is measured using our tools.

We propose a three-phase experiment workflow to evaluate the performance of blockchain as follows.
\begin{enumerate}
\item Create accounts, mint for them and store them in the account pool. Each account has enough balance and has access to all global states on blockchain.
\item Generate several clients and run these clients concurrently. Each client is assigned with different accounts from the pool and submits transactions using these accounts. 
\item Periodically send requests to the blockchain to query the progress of the transactions. 
\end{enumerate} 

The configuration of each client consists of two parts, 
the number of transactions and the content of transactions. 
By changing the number of transactions, we can simulate different workloads to evaluate the performance and robustness of these two platforms. 
Except for evaluating the whole platform, 
we can also focus on a specific layer and ignore the influence of other layers by changing the content of transactions.
Specifically, in the experiment, we conduct three types of transaction scripts to evaluate different components respectively, \ie the whole platform, the consensus layer, and the execution layer. 

\subsubsection{P2P Transfer Script}

As a global cryptocurrency, the most commonly used transaction script of Libra is P2P Transfer Script, thus we choose it to evaluate the whole platform. 
The transfer script is shown in Fig.~\ref{code:p2ptransfer},
which takes the address of \textit{payee} and transfer \textit{amount} as parameters, 
and calls the \textit{LibraAccount.pay\_from\_sender} function to transfer.

\begin{figure}[htbp]
	\begin{lstlisting}[language=Solidity]
  import 0x0.LibraAccount;
  main (payee: address, amount: u64) {
    LibraAccount.pay_from_sender(move(payee), move(amount));
    return;
  }
	\end{lstlisting}
	\caption{P2P Transfer Script}
	\label{code:p2ptransfer}
\end{figure}

\subsubsection{Do-nothing Script}

As shown in Fig.~\ref{code:donothing}, we employ Do-nothing Script to evaluate the performance of the consensus layer.
This script has the simplest logic and its execution has almost no impact on the performance of the whole platform. Besides, the execution of this script will not trigger any events or changes in account status, which can make the cost of storage also reach the minimum.
Therefore, by sending this transaction script, the performance of blockchain will mainly depend on communication for consensus, and the evaluation results can represent the performance of the consensus layer.


\begin{figure}[htbp]
	\begin{lstlisting}[language=Solidity]
  main() {
    return;
  }
	\end{lstlisting}
	\caption{Do-nothing Script}
	\label{code:donothing}
\end{figure}

\subsubsection{VM-heavy Script}

VM-heavy Script shown in Fig.~\ref{code:vmheavy} implements a loop to calculate the Fibonacci sequence in Move. This script needs lots of time to execute. 
In this part, we only change the number of iterations so the difference of performance under these situations can be used to evaluate the execution layer. 
In our experiment, the number of iterations varies from 1 to 1000 to evaluate the robustness and performance of VM.

\begin{figure}[htbp]
	\begin{lstlisting}[language=Solidity]
main() {
    let i: u64;
    let x: u64;
    let y: u64;
    let z: u64;
    let tmp: u64;
    i = 0;
    x = 1;
    y = 1;
    z = 2;
    while (copy(i) < 1000) { 
    //The number of iterations can be changed to simulate different complexity. 
        i = copy(i) + 1;
        tmp = copy(z);
        z = copy(x)+copy(y); 
        x=copy(y);
        y=copy(tmp);
    }

    return;
}
	\end{lstlisting}
	\caption{VM-heavy Script}
	\label{code:vmheavy}
\end{figure}

\subsection{Evaluation Metrics}
We use two metrics to evaluate the performance of blockchain platforms:
\begin{itemize}
     \item Tps: It is the number of transactions successfully processed and committed by the blockchain per second. This index reflects the throughput of the platform.
     \item Execution Time: It is the time consumed by the blockchain platform to process and commit a batch of transactions.   
\end{itemize}

The scalability of different blockchain platforms can be evaluated by changing the number of peers and observing the differences of Tps and execution time.



In our experiment, We have $n$ clients, each client has a sender account and a receiver account. For the $i$-th client, it firstly
sends a certain number of transactions from the sender account to the receiver account 
to make the blockchain \emph{warm up}. 
Then we make the first query to get the sequence number of the sender as $seq_i^{s}$ and record the local time as $t_i^{s}$. After that, we send all left transactions, make the second query and get the sequence number of the sender as $seq_i^{e}$ and record the local time as $t_i^{e}$. Then we calculate Tps as follow:

\begin{equation}
Average~Tps = \frac{\sum_{i}^{n}\frac{seq_i^{e}-seq_i^{s}}{t_i^{e}-t_{i}^{s}}} {n}
\end{equation}

\subsection{Infrastructure Setup and Configuration}
The infrastructure that the experiments are conducted on a physical server with 2 Intel E5-2680 v4 CPU (14 cores and 28 threads each), 384G RAM and Ubuntu 18.04 as OS.

For Fabric, we use the docker and docker-compose to deploy our own private blockchain. Both of them are tools provided by Hyperledger Fabric. The version of Fabric is 1.4.1.

For Libra, we compile it in the release mode and use the \emph{Libra\_swarm} modular to deploy our own private blockchain in a LAN.
Each validator in the blockchain network listens on an independent \emph{Access Control} port and communicates with each other through network.  

Each evaluation process corresponds to 12 concurrent clients which will be used to send transactions and queries. All results we obtained is the average value of at least 5 independent experiments.

\section{Results and Discussion}
\label{sec:results}
\begin{figure*}
	\centering
	\begin{subfigure}{0.42\linewidth}
		\centering
		\includegraphics[width=\linewidth]{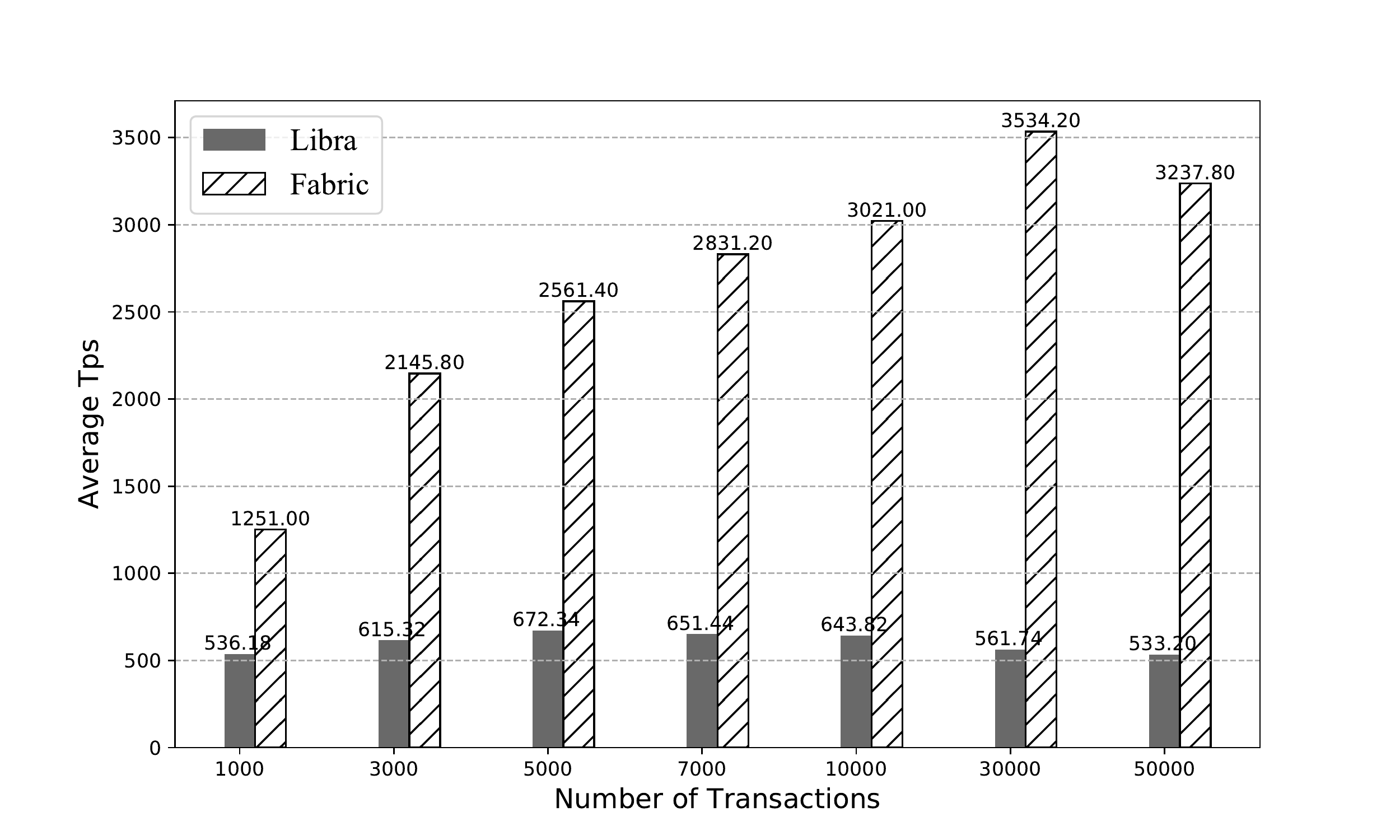}
		\caption{Average Tps with varying number of transactions}
		\label{fig:tpstxn}
	\end{subfigure}
	\quad
	\begin{subfigure}{0.42\linewidth}
		\centering
		\includegraphics[width=\linewidth]{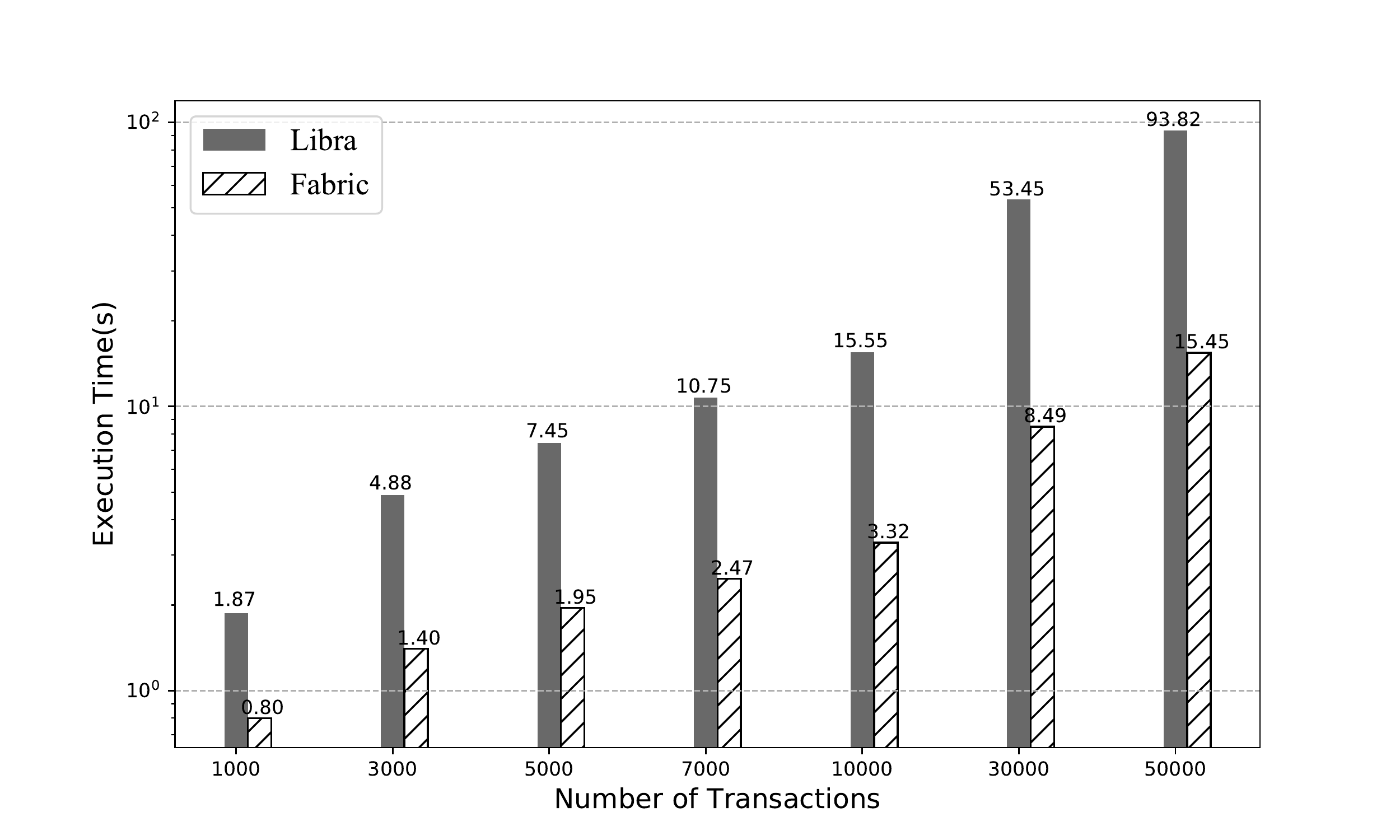}
		\caption{Execution Time with varying number of transactions}
		\label{fig:LatencyTxn}
	\quad
	\end{subfigure}
	\begin{subfigure}{0.42\linewidth}
		\centering
		\includegraphics[width=\linewidth]{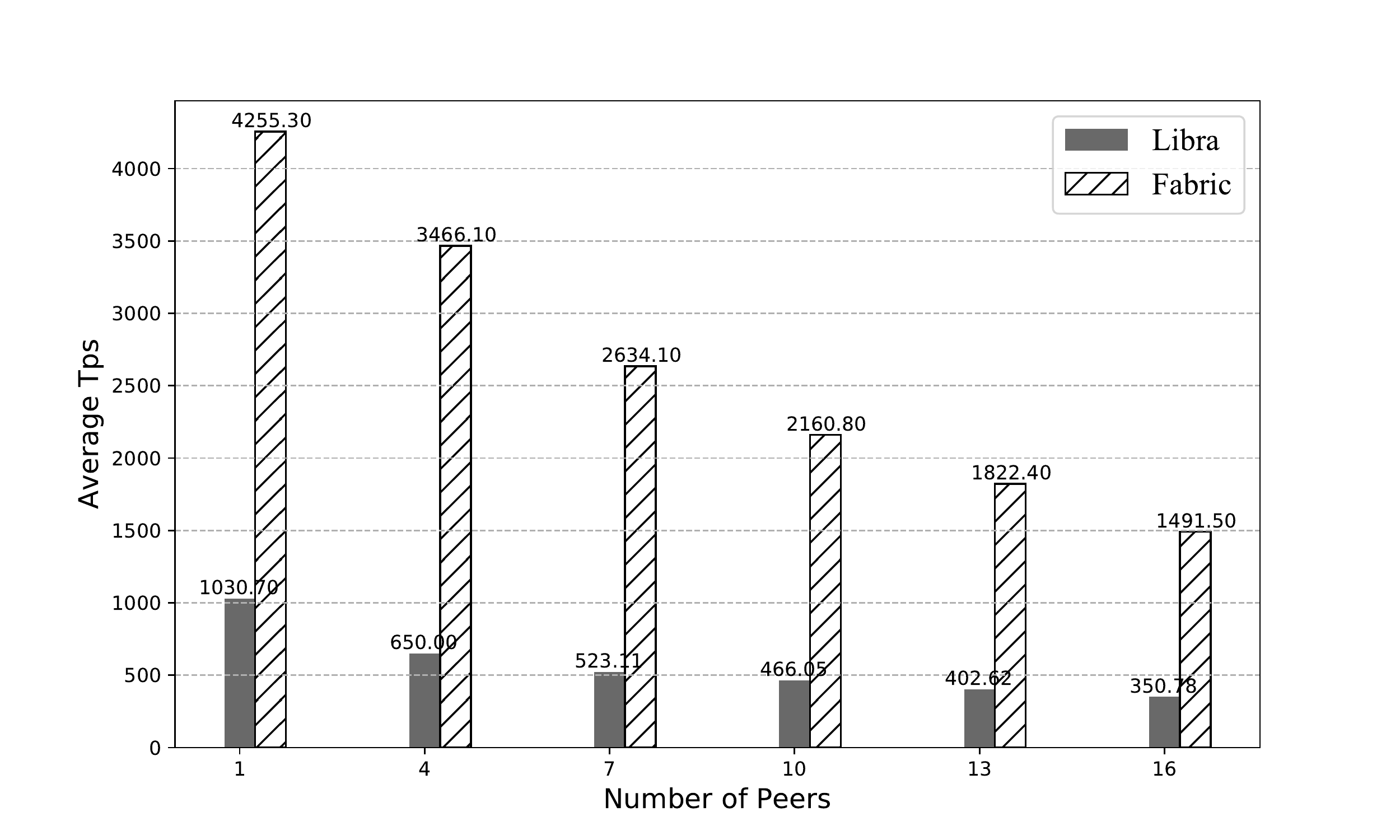}
		\caption{Average Tps with varying number of peers}
		\label{fig:tpspeer}
	\end{subfigure}
	\quad
	\begin{subfigure}{0.42\linewidth}
		\centering
		\includegraphics[width=\linewidth]{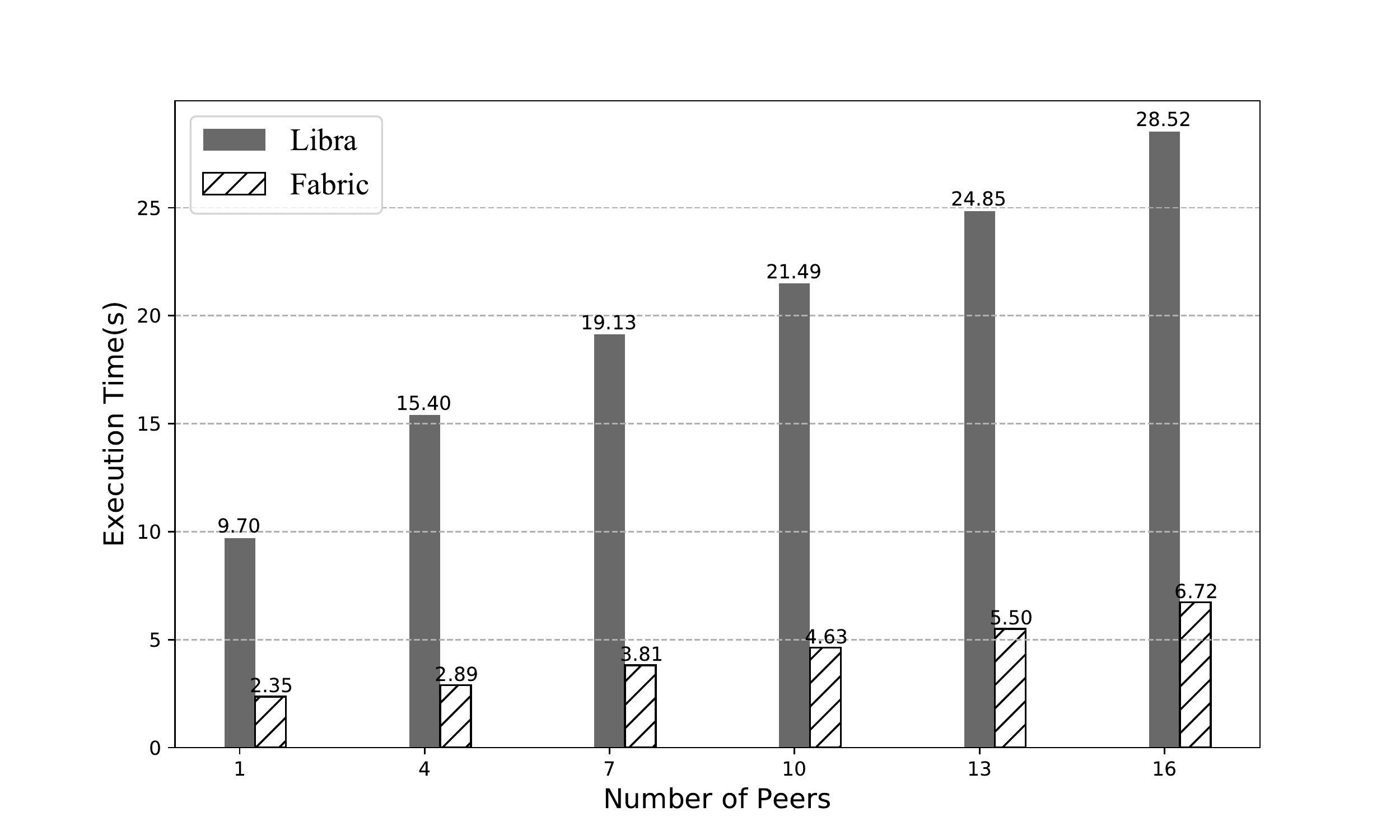}
		\caption{Execution Time with varying number of peers}
		\label{fig:latencypeer}
	\end{subfigure}
	\caption{Performance and Scalability of Libra and Hyperledger Fabric}

\end{figure*}

Our experiment consists of three parts:
\begin{itemize}
	\item Evaluate the performance of Libra using Fabric as the benchmark;
	\item Evaluate the scalability of Libra using Fabric as the benchmark;
	\item Evaluate the influence of different layers of Libra to find the bottlenecks and discuss them for better understanding and feasible improvements.
\end{itemize}

\subsection{Performance Evaluation}
In this part, we assign the number of peers to be 4 because a blockchain network with 4 peers can honestly complete the whole process composed by validation, consensus, execution, and storage. The transaction script we submit is P2P Transfer Script. We evaluate Libra and Fabric against different workloads by changing the number of transactions. 

\subsubsection{Comparing Average Throughput}
Fig.~\ref{fig:tpstxn} shows the throughput of these two platforms with different workloads. 
Both of them increase first and then decrease when the number of transactions increases. 
Libra's Tps reaches the maximum value when the number of transactions is about 5000 while 
Fabric's Tps reaches the maximum value when the number of transactions is about 30000. 
It shows Fabric has support to heavier workloads and has a better performance than Libra.

\subsubsection{Comparing Execution Time}
Fig.~\ref{fig:LatencyTxn} shows the execution time of those two platforms against different workloads. Both of them increase when the number of transactions increases. 

\subsection{Scalability Evaluation}

\begin{figure*}
	\centering
	\begin{subfigure}{0.42\linewidth}
		\centering
		\includegraphics[width=\linewidth]{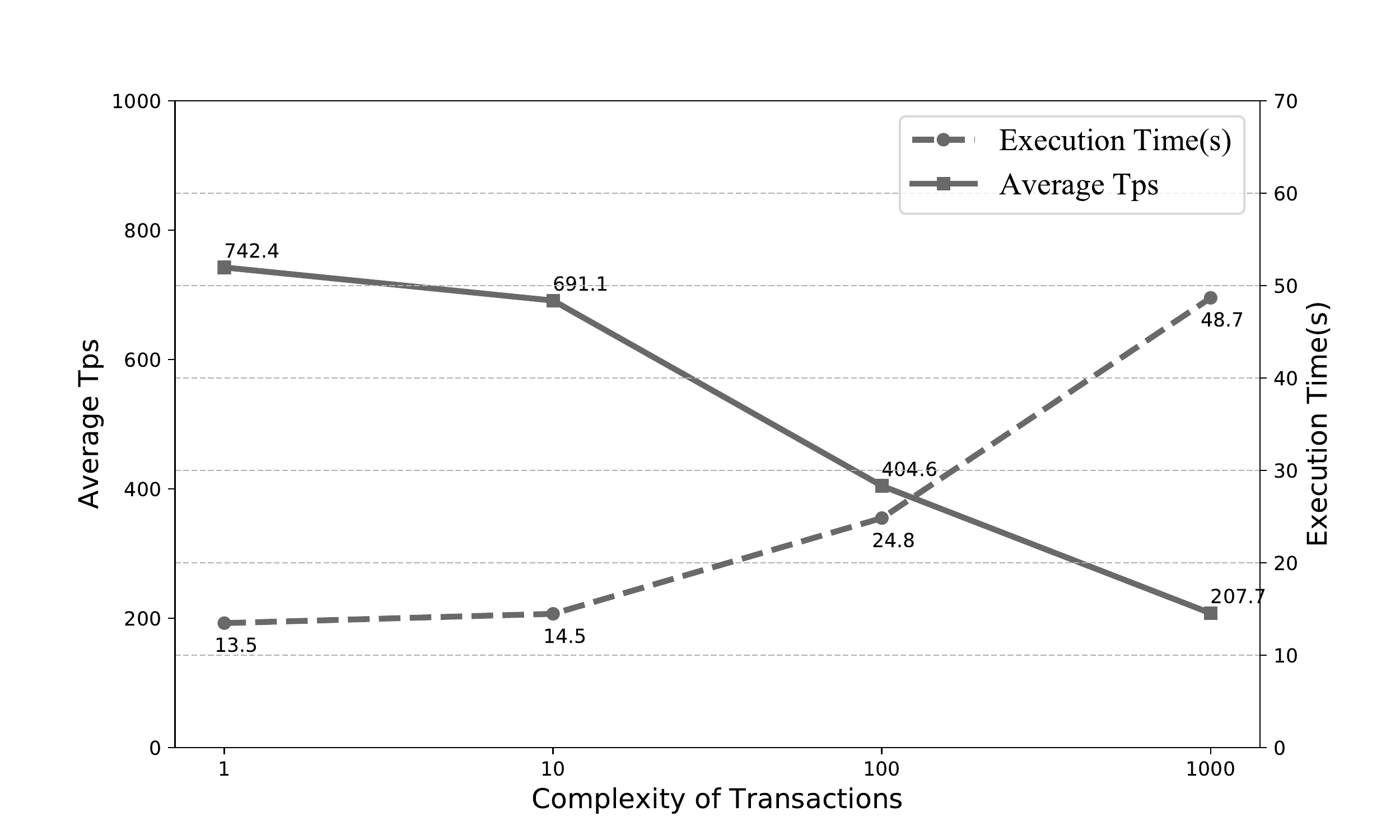}
		\caption{Performance of the execution layer}
		\label{fig:exetps}
	\end{subfigure}
	\begin{subfigure}{0.42\linewidth}
		\centering
		\includegraphics[width=\linewidth]{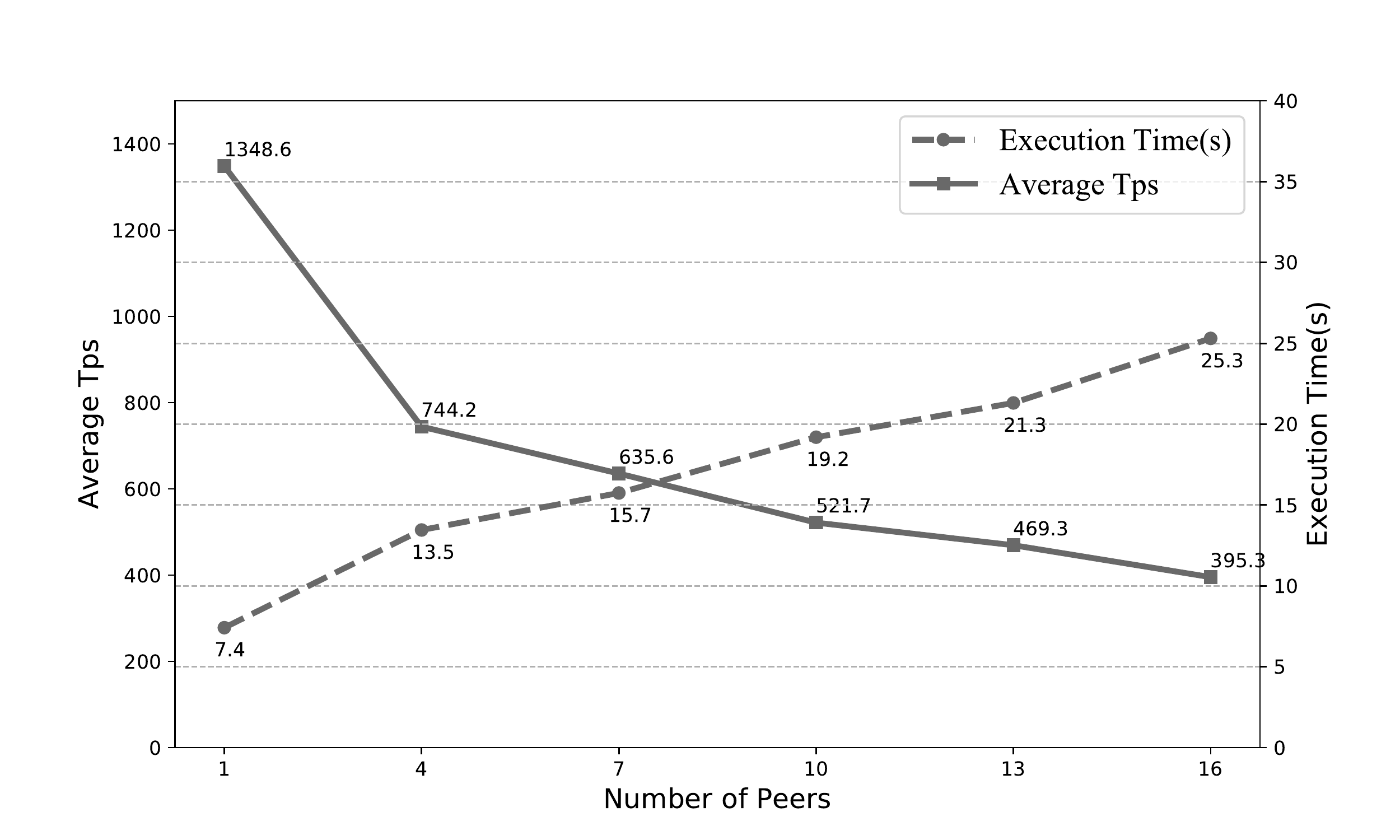}
		\caption{Performance of the consensus layer}
		\label{fig:consensus}
	\end{subfigure}
	\caption{Performance of different layers of Libra}
	
\end{figure*}

\begin{table*}
	\centering
	\caption{Metrics Std.}
	\label{tab:std}
	\scalebox{1.2}{
	\begin{tabular}{|c|c|c|c|c|c|c|c|c|}
		\hline
		\multirow{3}{*}{Performance} & Transactions & 1000  & 3000   & 5000   & 7000  & 10000  & 30000 & 50000 \\ \cline{2-9} 
		& Fabric's std & 26.72 & 144.91 & 54.16  & 64.26 & 119.78 & 64.67 & 49.9  \\ \cline{2-9} 
		& Libra's std  & 26.59 & 25.70  & 30.48  & 16.04 & 21.08  & 15.49 & 11.56 \\ \hline
		\multirow{3}{*}{Scalability} & Peers        & 1     & 4      & 7      & 10    & 13     & 16    & ---   \\ \cline{2-9} 
		& Fabric's std & 184.66 & 154.58 & 168.58 & 75.61 & 82.08  & 79.79 & ---   \\ \cline{2-9} 
		& Libra's std  & 17.24 & 19.55  & 12.39  & 18.49 & 9.51   & 8.32  & ---   \\ \hline
		\multirow{2}{*}{Consensus}   & Peer         & 1     & 4      & 7      & 10    & 13     & 16    & ---   \\ \cline{2-9}  
		& Libra's std  & 11.33 & 28.68  & 17.73  & 19.80 & 9.07   & 7.56  & ---   \\ \hline
		\multirow{2}{*}{Execution}   & Complexity   & 1     & 10     & 100    & 1000   & ---    & ---    & ---   \\ \cline{2-9}  
		& Libra's std  & 8.14 & 17.69  & 29.10  & 21.36& ---   & ---  & ---   \\ \hline
	\end{tabular}
	}
\end{table*}

In this part, we evaluate the scalability of Fabric and Libra with the number of peers assigned to be 1, 4, 7, 10, 13, 16 respectively. The number of transactions is fixed to be 10000 and the content of them is P2P Transfer Script. 
The result shows LibraBFT has good scalability.
 
\subsubsection{Comparing Average Throughput}

Fig.~\ref{fig:tpspeer} shows the average throughput of two platforms with the number of peers 
assigned to be 1, 4, 7, 10, 13, 16 respectively. 
The number of transactions is fixed to be 10000 to simulate a heavy workload. The throughput of both platforms decreases when the number of peers increases. When the number of peers increases from 1 to 16, the Tps of Fabric reduces to 35.1\% of the original while Libra's Tps reduces to 33.9\%. 
It represents that Libra has a high scalability as a BFT-based protocols. since the CFT-based consensus protocols, \ie Raft used by Fabric, usually has a better scalability than BFT-based protocols \cite{bach2018comparative}. However, that is still not enough. 
Libra's Tps is about 350 when the number of peers is 16. From the trend of Tps, we can predict that if the number of peers reaches 100 in the future as stated in the white paper \cite{Libra2019}, the performance of it will be greatly reduced and it will be difficult for Libra to provide global services.

\subsubsection{Comparing Execution Time}

Fig.~\ref{fig:latencypeer} shows the execution time of two platforms with different number of peers. 
Both of them increase when the number of peers increases. As the number of transactions is fixed to be 10000 and content of transactions is fixed to be P2P Transfer Script, we can draw the conclusion that the increase of the execution time is caused by the increase of the peers. It can present the impact of the number of peers on the time spent on consensus. For Libra, when the number of peers changed from 1 to 4, the execution time increased by 5.7s. After that, for every three more peers, the execution time increased at least 2.36s. For Fabric, the number is about 0.8s for every three more peers.

\subsection{Influence of Different Layers}
\subsubsection{The Execution Layer}
In this part, we evaluate the performance of the execution layer of Libra by 
changing the complexity of the transaction script we submit. 
We use the number of iterations in VM-heavy Script to simulate different complexity of transactions. 
The number of peers is fixed to be 4 and the number of transactions is fixed to be 10000.

Fig.~\ref{fig:exetps} shows the performance of the execution layer of Libra. 
By calculating Fibonacci Sequence iteratively, we evaluate the efficiency of the control flow and basic operations of Move VM. 
The result shows that with the exponential growth of complexity, the Tps of Libra decreases in a much slower way. 

\subsubsection{The Consensus Layer}
In this part, we evaluate the consensus layer of Libra by sending Do-nothing Script. The number of transactions is fixed to be 10000. The number of peers is assigned to be 1, 4, 7, 10, 13, 16 respectively

Fig.~\ref{fig:consensus} shows the performance of the consensus layer of Libra with varying number of peers 
. The throughput is a little higher than the above results shown in Fig.~\ref{fig:tpspeer} because the transactions we submit contains Do-nothing Script rather than P2P Transfer Script so that the execution and storage costs are lower. The overall trend is basically the same as Fig.~\ref{fig:tpspeer}.
\subsection{Standard derivation}

Table.~\ref{tab:std} shows the standard derivation of the Tps shown in the above figures. The standard derivation is calculated from at least 5 independent experiments. All standard derivations are in a reasonable range, which means those platforms have stable performance and also ensure the accuracy and correctness of our results.

\subsection{Further Discussion}

In the first two parts of our experiment, \ie the evaluation of the performance and scalability, we find that Fabric outperforms Libra in all occasions. This is mainly because Hyperledger Fabric v1.4.1 has a high code maturity and it has done a lot of work to improve its performance. Besides, it has a CFT-based consensus protocol which 
gives up byzantine fault tolerance in exchange for performance. 

To understand Libra better for further improvements, we selected 4 groups of experimental results, and listed them in Table~\ref{tab:control-variable}. contains three indexes and we explain them separately as follows. 

\begin{enumerate}
\item For the execution layer, \emph{on} means the transactions we choose to submit contains the P2P Transfer Script as it is the most common in actual use. And \emph{off} means the transaction contains the Do-nothing Script. If we choose \emph{off}, the time of execution and storage will reach the minimum and we regard that as splitting the execution layer out. 
\item For the consensus layer, \emph{on} means there are 4 peers in the blockchain network and they need to reach consensus for each transaction we submit while \emph{off} means there is only 1 peer in the blockchain so that the consensus can be reached without communicating with any other peers so that we can regard it as splitting the consensus layer out.
\item The other layers are necessary overhead related to the algorithm of the blockchain and the database. The states of them are always \emph{on}. 
\end{enumerate} 

\newcommand{\tabincell}[2]{\begin{tabular}{@{}#1@{}}#2\end{tabular}}

\begin{table}[]
	\centering
	\caption{Performance of the Libra blockchain under the combinations of different states of each layer}
	\label{tab:control-variable}
	\scalebox{1.2}{
	\begin{tabular}{|c|c|c|c|}
		\hline
		\tabincell{c}{Execution \\ Layer} & \tabincell{c}{Consensus \\ Layer} & \tabincell{c}{Other \\ Layers} & Tps

\\ \hline
		on                                   & on                                  & on                               & 643.82                   \\ \hline
		on                                   & off                                   & on                               & 1030.7                   \\ \hline
		off                                    & on                                  & on                               & 744.22                   \\ \hline
		off                                    & off                                   & on                               & 1348.68                  \\ \hline
	\end{tabular}
	}
\end{table}


Each transaction on the Libra blockchain is processed serially by different layers of the platform. 
In this way, with combinations of different states of those layers, the performance of the whole platform will change accordingly. The impact of the bottleneck layer's state on the overall performance should be greatest.

By comparing the first three rows with the 4-th row, we can see that with the consensus layer on and the execution layer off, the Tps of the whole platform decreases from 1348.68 to 744.22, when the consensus layer off and the execution layer on, the Tps of the whole platform decreases from 1348.68 to 1030.7. We can conclude that the consensus layer has a bigger impact on the whole performance, hence the bottleneck of Libra is the consensus layer, \ie LibraBFT. Besides, Table~\ref{tab:control-variable} can also measure the impact of different layers on the whole platform.



\section{Related Work}
\label{sec:relatedwork}
Due to the importance of performance and scalability, there has been a lot of previous work focusing on those metrics of different blockchain networks \cite{baliga2018performance} \cite{sukhwani2018performance}. They usually choose Hyperledger Fabric and Ethereum as targets, use Tps as their metrics, and change the number of peers to measure the scalability of blockchain networks.

Vukoli{\'c} \etal \cite{vukolic2015quest} compared PoW-based blockchains with BFT-based blockchains. The results showed that the performance of BFT-based blockchains is better while the PoW-based blockchains have better scalability. Harish Sukhwani \etal \cite{sukhwani2017performance} used Stochastic reward nets to model the PBFT consensus process. Their model included three main steps in PBFT consensus process, \ie Pre-Prepare, Prepare and Commit. They found PBFT is the bottleneck of Hyperledger Fabric v0.6 when the number of peers becomes large. LibraBFT is a new protocol based on HotStuff\cite{yin2019hotstuff}. As an extension to their work, we focus on different consortium blockchains and compare LibraBFT with a common consensus mechanism used by Fabric v1.4.1, \ie Raft. 

Dinh \etal \cite{dinh2018untangling} proposed a benchmarking framework called BLOCKBENCH and used it to analyze the performance of three major private blockchains, \ie Ethereum, Hyperledger Fabric v0.6 and Parity. They split those blockchains into four concrete layers and evaluating each layer against different workloads. By doing this, they find that the main bottleneck in Hyperledger Fabric v0.6 and Ethereum is the consensus protocol. The also found the execution engine of Ethereum is less efficient than Hyperledger's. Our work uses similar method, but we choose two new platforms which haven't been evaluated and compared and we are only interested in the consensus layer and the execution layer as Move VM and LibraBFT are main new features Libra proposed. Nasir \cite{nasir2018performance} proposed an evaluation framework architecture to compare the performance of the two different versions of Hyperledger Fabric (v0.6 and v1.0) against different workloads and numbers of peers. Thakkar \etal \cite{thakkar2018performance} observed the impact of various configuration parameters and then proposed three optimizations which improved the overall throughput of Fabric by 16x. 

Suankaewmanee \etal \cite{suankaewmanee2018performance} proposed a new m-commerce system model called MobiChain to apply blockchains technology to m-commerce. They conducted real experiments to evaluate the performance of the system. The result showed applying blockchain to m-commerce is an efficient way. In fact, Libra\cite{Libra2019} is also proposed for financial services innovation. So our work evaluates its m-commerce features and metrics as a global currency. Amsden \etal \cite{Amsden2019} introduce the main parts of Libra and they use approximate analysis to show that the Tps of Libra is likely to meet 1000. Their work is a theoretical and rough conjecture without experiments. So we conduct our tools to finish the study with credible result. 


\section{Conclusion}
\label{sec:conclusion}
In this paper, we propose a three-phase methodology to evaluate the performance and scalability of blockchain platforms.
We have implemented our methodology on the Libra blockchain and Hyperledger Fabric to explore their properties.
The experimental results show that the Libra blockchain still needs further improvements to meet the needs of global currency.
Moreover, We evaluated different layers of the Libra blockchain respectively to identify the bottleneck, help the community understand better and propose solutions to overcome the limitations of Libra.
In the future, we plan to follow up updates of the Libra blockchain and evaluate its performance.

\section*{Acknowledgment}
\label{sec:acknowledgement}
This work is supported by National Natural Science Foundation of China under the grant no. 61672060.
We also wish to thank Lifeng Ren, Xiangjun Feng and Shike Liu from Boya Blockchain Inc. for their helpful feedback and discussion.

\bibliographystyle{IEEEtran}
\bibliography{libra}

\begin{thebibliography}{10}
\providecommand{\url}[1]{#1}
\csname url@samestyle\endcsname
\providecommand{\newblock}{\relax}
\providecommand{\bibinfo}[2]{#2}
\providecommand{\BIBentrySTDinterwordspacing}{\spaceskip=0pt\relax}
\providecommand{\BIBentryALTinterwordstretchfactor}{4}
\providecommand{\BIBentryALTinterwordspacing}{\spaceskip=\fontdimen2\font plus
\BIBentryALTinterwordstretchfactor\fontdimen3\font minus
  \fontdimen4\font\relax}
\providecommand{\BIBforeignlanguage}[2]{{%
\expandafter\ifx\csname l@#1\endcsname\relax
\typeout{** WARNING: IEEEtran.bst: No hyphenation pattern has been}%
\typeout{** loaded for the language `#1'. Using the pattern for}%
\typeout{** the default language instead.}%
\else
\language=\csname l@#1\endcsname
\fi
#2}}
\providecommand{\BIBdecl}{\relax}
\BIBdecl

\bibitem{nakamoto2008bitcoin}
S.~Nakamoto \emph{et~al.}, ``Bitcoin: A peer-to-peer electronic cash system,''
  2008.

\bibitem{vukolic2017rethinking}
M.~Vukoli{\'c}, ``Rethinking permissioned blockchains,'' in \emph{Proceedings
  of the ACM Workshop on Blockchain, Cryptocurrencies and Contracts}.\hskip 1em
  plus 0.5em minus 0.4em\relax ACM, 2017, pp. 3--7.

\bibitem{wood2014ethereum}
G.~Wood \emph{et~al.}, ``Ethereum: A secure decentralised generalised
  transaction ledger,'' \emph{Ethereum project yellow paper}, vol. 151, no.
  2014, pp. 1--32, 2014.

\bibitem{garay2015bitcoin}
J.~Garay, A.~Kiayias, and N.~Leonardos, ``The bitcoin backbone protocol:
  Analysis and applications,'' in \emph{Annual International Conference on the
  Theory and Applications of Cryptographic Techniques}.\hskip 1em plus 0.5em
  minus 0.4em\relax Springer, 2015, pp. 281--310.

\bibitem{Lee:2019:BPR:3307334.3328666}
\BIBentryALTinterwordspacing
D.~R. Lee, Y.~Jang, H.~Jang, and H.~Kim, ``80
  enough - towards secure and efficient pow-based blockchain consensus
  (poster),'' in \emph{Proceedings of the 17th Annual International Conference
  on Mobile Systems, Applications, and Services}, ser. MobiSys '19.\hskip 1em
  plus 0.5em minus 0.4em\relax New York, NY, USA: ACM, 2019, pp. 639--640.
  [Online]. Available: \url{http://doi.acm.org/10.1145/3307334.3328666}
\BIBentrySTDinterwordspacing

\bibitem{Androulaki:2018:HFD:3190508.3190538}
\BIBentryALTinterwordspacing
E.~Androulaki, A.~Barger, V.~Bortnikov, C.~Cachin, K.~Christidis, A.~De~Caro,
  D.~Enyeart, C.~Ferris, G.~Laventman, Y.~Manevich, S.~Muralidharan, C.~Murthy,
  B.~Nguyen, M.~Sethi, G.~Singh, K.~Smith, A.~Sorniotti, C.~Stathakopoulou,
  M.~Vukoli\'{c}, S.~W. Cocco, and J.~Yellick, ``Hyperledger fabric: A
  distributed operating system for permissioned blockchains,'' in
  \emph{Proceedings of the Thirteenth EuroSys Conference}, ser. EuroSys
  '18.\hskip 1em plus 0.5em minus 0.4em\relax New York, NY, USA: ACM, 2018, pp.
  30:1--30:15. [Online]. Available:
  \url{http://doi.acm.org/10.1145/3190508.3190538}
\BIBentrySTDinterwordspacing

\bibitem{castro1999practical}
M.~Castro, B.~Liskov \emph{et~al.}, ``Practical byzantine fault tolerance,'' in
  \emph{OSDI}, vol.~99, no. 1999, 1999, pp. 173--186.

\bibitem{bach2018comparative}
L.~Bach, B.~Mihaljevic, and M.~Zagar, ``Comparative analysis of blockchain
  consensus algorithms,'' in \emph{2018 41st International Convention on
  Information and Communication Technology, Electronics and Microelectronics
  (MIPRO)}.\hskip 1em plus 0.5em minus 0.4em\relax IEEE, 2018, pp. 1545--1550.

\bibitem{lewis2017blockchain}
R.~Lewis, J.~McPartland, and R.~Ranjan, ``Blockchain and financial market
  innovation,'' \emph{Economic Perspectives}, vol.~41, no.~7, pp. 1--17, 2017.

\bibitem{Amsden2019}
\BIBentryALTinterwordspacing
Z.~Amsden, R.~Arora, S.~Bano, M.~Baudet, S.~Blackshear, A.~Bothra, and
  G.~Cabrera, ``{The Libra Blockchain},'' pp. 1--29, 2019. [Online]. Available:
  \url{https://developers.libra.org/docs/the-libra-blockchain-paper}
\BIBentrySTDinterwordspacing

\bibitem{ongaro2014search}
D.~Ongaro and J.~Ousterhout, ``In search of an understandable consensus
  algorithm,'' in \emph{2014 $\{$USENIX$\}$ Annual Technical Conference
  ($\{$USENIX$\}$$\{$ATC$\}$ 14)}, 2014, pp. 305--319.

\bibitem{Blackshear2019}
S.~Blackshear, E.~Cheng, D.~L. Dill, V.~Gao, B.~Maurer, T.~Nowacki, A.~Pott,
  S.~Qadeer, D.~Russi, S.~Sezer, T.~Zakian, and R.~Zhou, ``{Move : A Language
  With Programmable Resources},'' pp. 1--26, 2019.

\bibitem{Baudet2019}
\BIBentryALTinterwordspacing
M.~Baudet, A.~Ching, A.~Chursin, G.~Danezis, F.~Garillot, Z.~Li, D.~Malkhi,
  O.~Naor, D.~Perelman, and A.~Sonnino, ``{State Machine Replication in the
  Libra Blockchain},'' pp. 1--41, 2019. [Online]. Available:
  \url{https://developers.libra.org/docs/state-machine-replication-paper}
\BIBentrySTDinterwordspacing

\bibitem{yin2019hotstuff}
M.~Yin, D.~Malkhi, M.~Reiterand, G.~G. Gueta, and I.~Abraham, ``Hotstuff: Bft
  consensus with linearity and responsiveness,'' in \emph{38th ACM symposium on
  Principles of Distributed Computing (PODC’19)}, 2019.

\bibitem{nasir2018performance}
Q.~Nasir, I.~A. Qasse, M.~Abu~Talib, and A.~B. Nassif, ``Performance analysis
  of hyperledger fabric platforms,'' \emph{Security and Communication
  Networks}, vol. 2018, 2018.

\bibitem{Libra2019}
\BIBentryALTinterwordspacing
Libra, ``{An Introduction to Libra},'' pp. 1--12, 2019. [Online]. Available:
  \url{https://libra.org/en-US/white-paper/}
\BIBentrySTDinterwordspacing

\bibitem{baliga2018performance}
A.~Baliga, N.~Solanki, S.~Verekar, A.~Pednekar, P.~Kamat, and S.~Chatterjee,
  ``Performance characterization of hyperledger fabric,'' in \emph{2018 Crypto
  Valley Conference on Blockchain Technology (CVCBT)}.\hskip 1em plus 0.5em
  minus 0.4em\relax IEEE, 2018, pp. 65--74.

\bibitem{sukhwani2018performance}
H.~Sukhwani, N.~Wang, K.~S. Trivedi, and A.~Rindos, ``Performance modeling of
  hyperledger fabric (permissioned blockchain network),'' in \emph{2018 IEEE
  17th International Symposium on Network Computing and Applications
  (NCA)}.\hskip 1em plus 0.5em minus 0.4em\relax IEEE, 2018, pp. 1--8.

\bibitem{vukolic2015quest}
M.~Vukoli{\'c}, ``The quest for scalable blockchain fabric: Proof-of-work vs.
  bft replication,'' in \emph{International workshop on open problems in
  network security}.\hskip 1em plus 0.5em minus 0.4em\relax Springer, 2015, pp.
  112--125.

\bibitem{sukhwani2017performance}
H.~Sukhwani, J.~M. Mart{\'\i}nez, X.~Chang, K.~S. Trivedi, and A.~Rindos,
  ``Performance modeling of pbft consensus process for permissioned blockchain
  network (hyperledger fabric),'' in \emph{2017 IEEE 36th Symposium on Reliable
  Distributed Systems (SRDS)}.\hskip 1em plus 0.5em minus 0.4em\relax IEEE,
  2017, pp. 253--255.

\bibitem{dinh2018untangling}
T.~T.~A. Dinh, R.~Liu, M.~Zhang, G.~Chen, B.~C. Ooi, and J.~Wang, ``Untangling
  blockchain: A data processing view of blockchain systems,'' \emph{IEEE
  Transactions on Knowledge and Data Engineering}, vol.~30, no.~7, pp.
  1366--1385, 2018.

\bibitem{thakkar2018performance}
P.~Thakkar, S.~Nathan, and B.~Viswanathan, ``Performance benchmarking and
  optimizing hyperledger fabric blockchain platform,'' in \emph{2018 IEEE 26th
  International Symposium on Modeling, Analysis, and Simulation of Computer and
  Telecommunication Systems (MASCOTS)}.\hskip 1em plus 0.5em minus 0.4em\relax
  IEEE, 2018, pp. 264--276.

\bibitem{suankaewmanee2018performance}
K.~Suankaewmanee, D.~T. Hoang, D.~Niyato, S.~Sawadsitang, P.~Wang, and Z.~Han,
  ``Performance analysis and application of mobile blockchain,'' in \emph{2018
  international conference on computing, networking and communications
  (ICNC)}.\hskip 1em plus 0.5em minus 0.4em\relax IEEE, 2018, pp. 642--646.

\end{thebibliography}

\end{document}